\documentclass[english,reprint,aps,superscriptaddress]{revtex4-2}
\usepackage[T1]{fontenc}
\usepackage[utf8]{inputenc}
\usepackage{babel}
\usepackage{mathrsfs}
\usepackage{bm}
\usepackage{amsmath}
\usepackage{mdframed}
\usepackage{tcolorbox}
\usepackage{amsthm}
\usepackage{amssymb}
\usepackage{lmodern}  
\usepackage{graphicx}
\usepackage[unicode=true,bookmarks=false,breaklinks=false,pdfborder={0 0 1},colorlinks=false]{hyperref}
\hypersetup{colorlinks,linkcolor=blue,citecolor=blue}
\makeatletter

\usepackage[normalem]{ulem}
\usepackage{babel}
\usepackage{verbatim}
\usepackage{amsthm}
\usepackage{xargs}[2008/03/08]
\usepackage{sidecap}

\newif\ifarxiv
\arxivtrue

\let\mainsection\section
\usepackage{color}%
\usepackage{newtxtext}

\ifarxiv
 \let\savedaddcontentsline\addcontentsline 
\global\long\def\addcontentsline#1#2#3{}%

\else

\usepackage{xr}

\newcommand*{\addFileDependency}[1]{%
\typeout{(#1)}%
\@addtofilelist{#1}
\IfFileExists{#1}{}{\typeout{No file #1.}}
}
\newcommand*{\myexternaldocument}[1]{%
\externaldocument{#1}%
\addFileDependency{#1.tex}%
\addFileDependency{#1.aux}%
}

\myexternaldocument{sm}

\fi

\definecolor{DarkBlue}{rgb}{0.0, 0.0, .7}

\newcommand\cut[1]{}

\global\long\def\papertitle{Neutral theory of cooperative dynamics}

\global\def\simp{\lambda}
\global\def\simpss{{\lambda^*}}

\global\def\simpnot{\lambda_\circ}

\global\def\mig{\mu}
\global\def\pp{P}
\global\def\n{\mathbf{n}}
\global\def\1{\boldsymbol{1}}

\global\def\erfcx{\mathsf{erfcx}}

\global\def\mrt{ \tau}
\global\def\mrtcore{\tau_{\mathsf{core}}}
\global\def\mrtout{ \tau_{\mathsf{out}}}

\global\def\Rcore{R^*_{\mathsf{core}}}
\global\def\Rout{R^*_{\mathsf{out}}}
\global\def\RFisher{R^*_{\mathsf{Logseries}}}

\global\def\DawsonOrig{\mathscr{D}}

\newcommand{\backward}[1]{d_{#1}}
\newcommand{\forward}[1]{b_{#1}}

\newcounter{myboxcounter}[section]

\newtcolorbox[auto counter, number within=section]{mybox}[2][]{colback=gray!20, colframe=white, fonttitle=\bfseries, title=Box~\thetcbcounter: #2,#1}

\allowdisplaybreaks

\makeatother

\begin{document}
\title{\papertitle}
\author{Jordi Pi\~nero}
\email{jpinerfe@gmail.com}
\thanks{Equal contribution.}
\affiliation{Department of Physics and Astronomy, Michigan State University, East Lansing, Michigan 48824, USA}
\author{Artemy Kolchinsky}
\email{artemyk@gmail.com}
\thanks{Equal contribution.}
\affiliation{ICREA-Complex Systems Lab, Universitat Pompeu Fabra, 08003 Barcelona,
Spain}
\affiliation{Barcelona Collaboratorium, Wellington 30, 08005 Barcelona, Spain}
\affiliation{Universal Biology Institute, University of Tokyo, 7-3-1 Hongo, Bunkyo-ku, Tokyo 113-0033, Japan}
\author{Sidney Redner}
\email{redner@santafe.edu}
\affiliation{Santa Fe Institute, Santa Fe, New Mexico 87501, United States}
\author{Ricard Solé}
\email{ricard.sole@upf.edu}
\thanks{Corresponding author.}
\affiliation{ICREA-Complex Systems Lab, Universitat Pompeu Fabra, 08003 Barcelona,
Spain}
\affiliation{Santa Fe Institute, Santa Fe, New Mexico 87501, United States}
\affiliation{Institut de Biologia Evolutiva (CSIC-UPF), 08003 Barcelona, Spain}
\begin{abstract}
  Mutualistic interactions are widespread in nature, from plant communities and microbiomes to human organizations. Along with competition for resources, cooperative interactions 
  shape biodiversity and contribute to the robustness of complex
  ecosystems. We present a stochastic neutral theory of cooperator species. Our model shares with the classic neutral theory of biodiversity the assumption that all species are equivalent, but crucially differs in requiring cooperation between species for replication.
  With low migration, our model displays a bimodal species-abundance distribution, with a high-abundance mode associated with a core of cooperating species. This core is responsible for maintaining a diverse pool of long-lived species, which are present even at very small migration rates. We derive analytical expressions of the steady-state species abundance distribution, as well as scaling laws for diversity, number of species, and residence times. 
  With high migration, our model recovers the results of classic neutral theory. 
  We briefly discuss implications of our analysis for research on the microbiome, synthetic biology, and the origin of life.
\end{abstract}
\maketitle

\mainsection{Introduction}
{C}ooperation is ubiquitous across scales in complex systems. In ecology, cooperative
interactions  
shape ecosystem structures~\cite{bronstein2015mutualism,leigh2010evolution}.
Within biology, mutualism (a reciprocal form of cooperation) is the engine of evolutionary transitions~\cite{lutzoni1997accelerated,schuster2001does} and constitutes an essential part of the architecture of biodiversity~\cite{bascompte2007plant,suweis2013emergence}. In microbiomes, cooperative interactions occur through extensive cross-feeding exchanges associated with shared diffusive metabolites~\cite{muller2014genetic,germerodt2016pervasive,culp2023cross}, a phenomenon dubbed the {\it social network} of microorganisms~\cite{zengler2018social}. Moreover, cooperative interactions enhance community stability~\cite{coyte2015ecology} and facilitate metabolic functions~\cite{morris2013microbial}. Cooperation has also been studied in human organizations, for example, between companies engaged in jointly manufacturing a certain product~\cite{saavedra2009simple}.

Ecological theory has traditionally studied either 
{\textit{(i)}} the dynamics of large randomly-assembled communities, leading to general stability-complexity principles~\cite{may2019stability,allesina2012stability}, or  {\textit{(ii)}} the stationary properties of stochastic interactions between neutral species, inspired by the neutral theory of biodiversity 
\cite{hubbell2011unified,sole2002self}.
Stability patterns in mutualistic communities under approach {\textit{(i)}} have been addressed using network models~\cite{bastolla2009architecture,valdovinos2019mutualistic,guimaraes2020structure}. The results of these models may depend strongly on the assumptions made about different species (in terms of size, lifestyle, or physiology) and their interactions, making it challenging to derive general theoretical lessons.

In contrast, approach {\textit{(ii)}} assumes that all species in the system are equivalent, shifting the focus to the effects of demographic noise as the main determinant of emerging ecological patterns~\cite{volkov2003neutral,harte2011maximum,azaele2016statistical,grilli2020macroecological,o2010field}. Despite its simplicity, neutral theory successfully accounts for many relevant statistical patterns, including abundance distributions, species-area relations, and diversity estimates in space and time~\cite{muneepeerakul2008neutral,hubbell2011unified,harte2011maximum,rosindell2011unified}. 
However, a neutral theory that explicitly involves cooperative interactions is lacking, and little is known about the impact of mutualistic exchanges under a neutral picture. For example, microbiomes exhibit marked quantitative patterns that diverge significantly from those predicted by classic neutral theory,  such as the presence of a persistent subset of species~\cite{morris2013microbial,neu2021defining,wu2024core}.

In this paper, we present a neutral theory for ecosystems of cooperators. Our model preserves the neutral hypothesis --- that all species in the system follow the same replication rules, albeit possibly with frequency-dependent replication probabilities. However, it departs from the classic neutral theory of biodiversity by requiring cooperation between species for replication to occur. Our model contains only two independent parameters: the total number of individuals in the system $N$, typically assumed to be large, and the migration rate $\mu$, the probability that a new species enters the system at a given time step. 
One of the remarkable features of our model is the emergence of a bimodal species-abundance distribution, characterized by a ``core'' of cooperator species that remain at high abundances for much longer compared to species in the low-abundance mode.  
This core leads to the maintenance of species diversity even at very small migration rates, where, in the absence of cooperation, the system would rapidly fixate. In this sense, we show that cooperative interactions can provide a powerful stabilizing effect. 

We solve our model analytically and confirm the results via extensive numerical simulations. Our analysis is split into two main parts: first, we focus on the statistical patterns displayed by the system's steady state; afterward, we consider the stochastic dynamics of abundance trajectories as species enter and leave the system.

\begin{figure}[t]
\begin{center}
\includegraphics[width=.7\columnwidth]{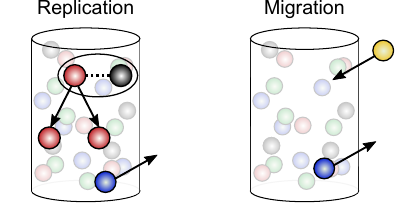}
\caption{\textbf{Neutral model of cooperators} The population evolves according to two rules: \textbf{(a)} during replication, individuals of different species (black and red balls) cooperate. The red individual replicates and replaces a randomly chosen individual (blue ball). \textbf{(b)} during migration with probability $\mig$, an individual from a new species (yellow ball)  enters from an external pool and substitutes a randomly-chosen individual (blue ball).}
\label{fig:1}
\end{center}
\end{figure}

\mainsection{Model setup}
Our model considers a well-mixed population of $N$ individuals, each labeled by a species.   
The configuration of the system at a given time is represented by the \emph{abundance vector} $\n=(n_1,n_2,\dots)$, where $n_i$ is the abundance of species $i$.  
We use $R(\n)$ to indicate the number of nonzero elements of vector $\n$, i.e., the number of present species. 
The configuration evolves stochastically over a sequence of discrete steps involving 
either a cooperative replication or a migration event:

\vspace{5pt}
\noindent 1) \emph{Replication with probability $1-\mig$, Fig.~\ref{fig:1} (left)}:  two individuals are randomly chosen from the population, individual $A$ from species $i$ and individual $B$ from species $j$. If $A$ and $B$ belong to the same species ($i=j$), nothing happens. Otherwise, an individual $C$ of species $k$ is randomly selected from the population and replaced by a new individual of species $i$. 
The abundance vector $\n$ is updated as
\begin{align}
    (n_i, n_k) \to (n_i + 1, n_k - 1)\,.\label{eq:replicator-event}
\end{align}
(If $i=k$, there is no change). Given our well-mixed population, the probability that this event occurs 
in a given step is
\begin{align}
(1-\mig) \frac{n_i}{N}\,\frac{N-n_i}{N} \,\frac{n_k}{N}.
\label{eq:probability-replication-event}
\end{align}
In \eqref{eq:probability-replication-event}, the factors represent the probability that: replication takes place, $A$ belongs to species $i$, $B$ belongs to any species other than $i$, and $C$ belongs to species $k$.

\vspace{5pt}
\noindent 2) \emph{Migration with probability $\mig$, Fig.~\ref{fig:1} (right)}:  an individual $A$ of species $i$ is randomly chosen and replaced by an individual of a new species. The abundance vector  $\n$ is updated as
\begin{align}
    (n_i, n_{R(\n)+1}) \to (n_i - 1, 1)\,.
    \label{eq:migration-event}
\end{align}
This event happens with probability $\mig(n_i/N)$ per step.
For notational convenience, if any species becomes extinct ($n_i \to 0$), the species are reindexed so that the first $R(\n)$ entries of $\n$ are strictly positive. 

Total population size is conserved under the above rules ($\sum_{i} n_i=N$), resulting in competition between species for limited space. Also, because every migration event introduces a new species, migration represents an inflow from an infinitely-diverse external reservoir.

In the following, we consider the stationary properties of this stochastic process.  We are particularly interested in the \emph{species abundance distribution} $\pp_n$ (the probability that a species has abundance $n$ in  steady state) and the system's overall diversity. We measure diversity using the Simpson index~\cite{simpson1949measurement}, defined for abundance vector $\n$ as\footnote{The Simpson index $\protect\simp(\n)$ is sometimes alternatively defined as ${\Huge \Sigma}_{i=1}^{S} \frac{n_i(n_i-1)}{N(N-1)}$, i.e., as the probability of drawing two individuals of the same species when sampling without replacement. The definition~\eqref{eq:simpsondef} corresponds to the same probability when sampling with replacement. The two definitions differ by a small term of order $1/(N-1)$, which is irrelevant for our analysis.}

\begin{align}
\simp(\n) :=
\sum_{i=1} \left(\frac{n_i}{N}\right)^2\,.
\label{eq:simpsondef}
\end{align}
The Simpson index is the probability that two randomly chosen individuals belong to the same species~\cite{simpson1949measurement}, and it is commonly used in ecology as an inverse measure of diversity~\cite{magurran2003measuring}.  It is bounded as $1/R(\n) \le \simp(\n) \le 1$, with the lower bound achieved when the population is evenly distributed among $R(\n)$ species (maximum diversity) and the upper bound achieved when the population is fixated on a single species (minimum diversity). The \emph{inverse Simpson index} $1/\simp(\n)$ may be interpreted as the ``effective'' number of species in the system.

\begin{figure}[t]
\begin{center}
\includegraphics[width=0.85\columnwidth]{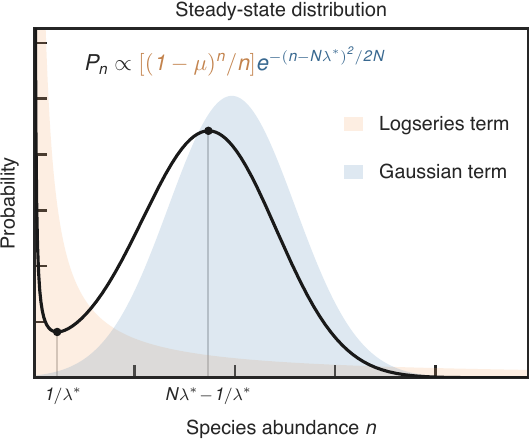}
\caption{\textbf{Schematic illustration of the species abundance distribution $P_n$} in steady state. This distribution depends on population size $N$, migration rate $\mig$, and the steady-state Simpson index $\simpss$. In Eq.~\eqref{eq:Pn1}, it is approximated as the  product of Logseries distribution, arising from competition for space, and a Gaussian distribution, arising from cooperative interactions.  At low migration, the combined distribution exhibits a bimodal shape, with local maxima at $n=1,n=N\simpss-1/\simpss$ and a local minimum at $n=1/\simpss$, see Eq.~\eqref{eq:nminmax}. Note that the two terms combine multiplicatively, not additively, so their areas do not add up to the area of the combined distribution.}
\label{fig:schematicss}
\end{center}
\end{figure}

As shown below, our model has two different steady-state regimes, depending on the migration rate.
When migration is high, diversity is very large ($\simp \approx 0 $), competition for space is strong, and the species abundance is well-described by the Fisher's Logseries distribution. This regime reduces to the classic neutral model of Hubbell, which does not involve cooperative interactions~\cite{hubbell1997unified}. 
When migration is low, the system forms a ``core'' of high-abundance species, and the species-abundance distribution acquires a characteristic bimodal shape, shown schematically in Fig.~\ref{fig:schematicss}. In what follows, we will study the emergence and nature of this core.

\mainsection{Steady-state distribution }
\label{sec:ss-dist}
To derive the steady-state species abundance distribution, we exploit the neutrality of our model, which allows us to treat all present species as statistically equivalent. 
Thus, without loss of generality, we consider species $i=1$ as the {\em representative species}.

The stochastic dynamics of the abundance of the representative species, written $n\equiv n_1$,  follows a birth-death process. 
The birth probability 
 ($n\to n + 1$ for $n\in\{1,\dots,N-1\}$) during a single time step is 
\begin{align}
\forward{n}&=(1-\mig)\frac{n}{N} \Big(1-\frac{n}{N}\Big)^2 \,,
\label{eq:birthprob}
\end{align}
while the death probability ($n\to n-1$ for $n\in\{1,\dots,N\}$) is
\begin{align}
\backward{n}&=(1-\mig)\frac{n}{N}\Big(1-\frac{n}{N}\Big)\Big[1-\Big(1-\frac{n}{N}\Big)\simpnot(\n)\Big]+\mig\frac{n}{N}\,.
\label{eq:deathprob}
\end{align}
See SM\ref{app:set-up} for a detailed derivation of Eqs.~\eqref{eq:birthprob}-\eqref{eq:deathprob}. The quantity $\simpnot(\n)$ refers to the Simpson index of the other species (i.e., excluding the representative species $i=1$):
\begin{align}
\simpnot(\n) :=\sum_{i=2} \left(\frac{n_i}{N-n}\right)^2 \,.
\label{eq:simpsonnotdef}
\end{align}
Although different species interact  during replication, 
the stochastic dynamics of each species depends on all others  only through a single number, $\simpnot(\n)$.  
We remark that similar coupled birth-death processes have been considered in work 
on interacting particle systems and nonlinear chemical master equations~\cite{malek-mansour_master_1975}.

The above birth-death process is absorbing into the extinction state $n=0$, since any {particular} species will eventually go extinct. To study the abundance of non-extinct species, we may make the process ergodic by adding a positive birth probability $b_0>0$ out of the extinction state $n=0$. In the following, we focus on the steady-state distribution restricted to positive abundances $n\in \{1,2,\dots\}$, which does not depend on the choice of $b_0$. 

We now find the steady-state species abundance distribution $\pp_n$ of the representative species. Since births and deaths must balance in steady state, $ \pp_n \forward{n}=\pp_{n+1} \backward{n+1} $ for $n\in\{1,\dots,N-1\}$, we have   
\begin{align}
P_{n}\propto \prod_{k=1}^{n-1} \frac{\forward{k}}{\backward{k+1}} \,.\label{eq:db}
\end{align}
This is not yet a closed equation because the death  probability~\eqref{eq:deathprob} depends on the fluctuating quantity  $\simpnot(\n)$, i.e., the  Simpson index of the non-representative species.  
However, when the number of species is large, 
$\simpnot(\n)\approx\lambda(\n)$. Futhermore, due to self-averaging, $\lambda(\n)$ 
is tightly peaked around its expected steady-state value,   
\begin{align}
\simpnot(\n) \approx \simpss :=\langle\simp(\n)\rangle,
\label{eq:ss-simpnot}
\end{align}
where $\langle \cdot \rangle$ indicates steady-state expectation over the entire population. 
Plugging~\eqref{eq:ss-simpnot} into~\eqref{eq:deathprob} and using~\eqref{eq:db} we find that, under reasonable assumptions, the steady-state abundance distribution is approximated as (see SM\ref{app:ss} for details):
\begin{align}
    \pp_n \propto \frac{\left(1-\mig\right)^{n}}{n}\, e^{-(n-N{\simpss})^2/2N} \,.
    \label{eq:Pn1}
\end{align}

\eqref{eq:Pn1} is one of our main results, showing how the abundance distribution depends on the population size $N$, migration rate $\mig$, and the steady-state Simpson index $\simpss$. In reality, $\simpss$ itself depends on $N$ and $\mig$ --- the only two parameters that describe the system --- although finding the explicit expression of $\simpss$ in terms of $N$ and $\mig$ is not trivial (we will do so in the next section). For now, we note that the Simpson index $\simpss$ and the migration probability $\mig$ move in opposite directions. When $\mig$ decreases to 0 (migration vanishes), $\simpss$ increases to 1 (fixation is reached). Conversely, when $\mig$ increases to 1 (only migration occurs), $\simpss$ decreases to its minimum value of $1/N$ (maximum diversity).

Observe that \eqref{eq:Pn1} expresses $\pp_n$ as the product of two distributions, as illustrated in Fig.~\ref{fig:schematicss}. The first is the Logseries distribution,  $\left(1-\mig\right)^{n}/n$, the abundance distribution of Hubbell's neutral model, which does not have  cooperative interactions~\cite{mckane2000mean}. This contribution represents migration-driven competition for space, and it dominates the system at high migration rates. The second is a Gaussian distribution, $e^{-(n-N\simpss)^2/2N}$, with mean abundance $N\simpss$ and standard deviation $\sqrt{N}$. This contribution represents a high-abundance  mode that emerges due to cooperative interactions. We term this contribution the \emph{cooperator core}.

The emergence of the cooperator core 
is a highly nontrivial phenomenon, and it implies  various other interesting aspects of this model (maintenance of high diversity at low migration, long residence times, etc.). In simple terms, the cooperator core arises due to the frequency-dependent (i.e., abundance-dependent)  replication rates. Individuals from low-abundance species are unlikely to be randomly paired with an individual from the same species, thus they have a high per-capita probability of replicating. 
Conversely, individuals from high-abundance species are more likely to be randomly paired with an individual from the same species, thus they have a lower per-capita replication probability. As a result, species feel an effective force towards an intermediate abundance at  $n\approx N\simpss$, leading to a bimodal abundance distribution as in Fig.~\ref{fig:schematicss}.

Importantly, the cooperative core only emerges at low migration rates. At high migration, when $\mig$ is large and $\simpss$ is low, most species exhibit low abundances near $n\approx 1$. In this regime, there is very little variation in species abundances, so the frequency dependence of the replication rates is negligible, and our system approaches Hubbell's neutral model. We also remark that bimodality requires that the mean of the Gaussian contribution ($N\simpss$) be many standard deviations ($ \sqrt{N}$) away from the origin or, in other words, that $\simpss\gg 1/\sqrt{N}$. Clearly, this cannot occur at very high migration probability, since $\simpss$ decreases to its minimum value of $1/N$ as $\mig$ increases to 1.  In the next section, we will derive the explicit migration probability $\mig$ that allows for bimodality.

\begin{figure*}[t]
\begin{center}
\includegraphics[height= 4.45cm]{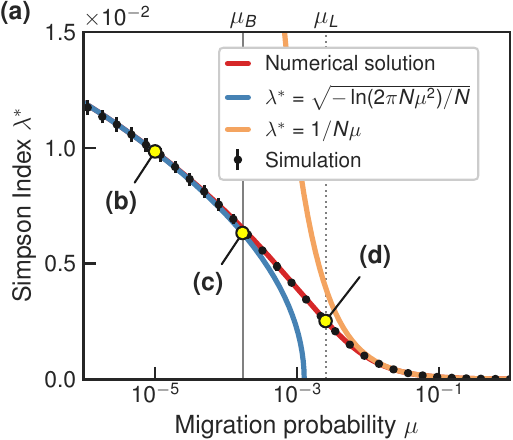}
\quad
\includegraphics[height= 4.45cm]{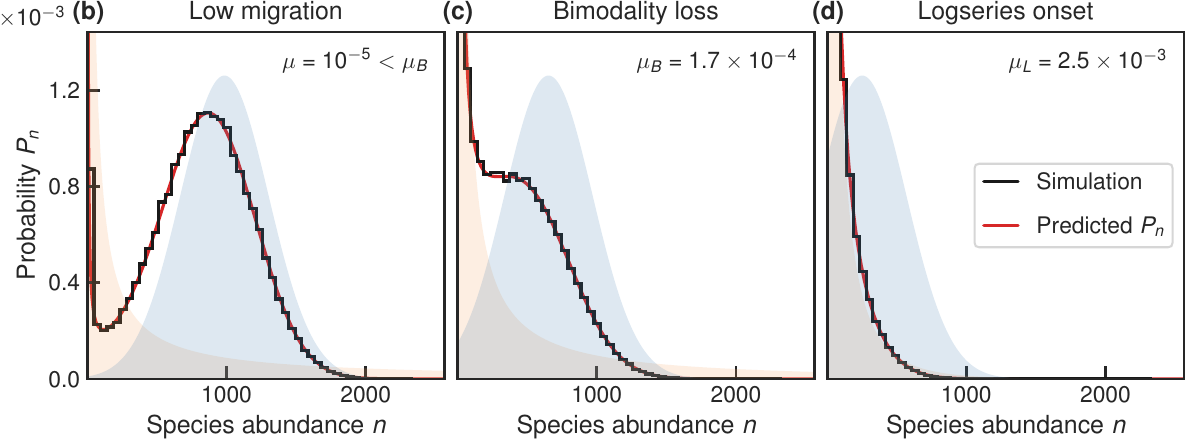}
\caption{\textbf{Scaling of Simpson index and species abundance distributions.} 
\textbf{(a)} Scaling of Simpson index versus migration probability $\mig$ in a system with $N=10^5$ individuals. We compare mean and standard deviations from simulations of the full system (100 runs; black), numerical inversion of Eq.~\eqref{eq:solA0} (red), approximations in low migration~\eqref{eq:ss-div} (blue) and high migration~\eqref{eq:simphighmig} (orange) regimes. Solid vertical line indicates $\mu_B$~\eqref{eq:bimodality} where bimodality is lost; dotted vertical line indicates  $\mu_L$~\eqref{eq:transition} where system transitions to Logseries regime. 
\textbf{(b)-(d)} Empirical histograms from simulations (across 1000 runs) versus predicted steady-state distributions~\eqref{eq:Pn1} for three migration probabilities. 
As in Fig.~\ref{fig:schematicss}, shaded areas represent Logseries (orange) 
and Gaussian (blue) contributions.
\textbf{(b)} Low-migration regime exhibits a bimodal distribution with a cooperator core. 
\textbf{(c)} Bimodality disappears once migration probability increases beyond $\mig_B$~\eqref{eq:bimodality}. 
\textbf{(d)} At higher migration probabilities, including the Logseries transition point where $\mig_L=\simpss$~\eqref{eq:transition}, the  distribution approaches the Logseries, as predicted by Hubbell's neutral theory.}
\label{fig:speciesabundance}
\end{center}
\end{figure*}

In this section, we derive several important properties of the steady-state species abundance distribution and the cooperator core. In particular, we derive expressions of the steady-state Simpson index, the characteristic migration probability $\mig_B$ below which the formation of the cooperator core takes place, and the expected number of species in steady state. Most of our analyses will consider separately the case of low migration (when the core is present) and high migration. 
Fig.~\ref{fig:speciesabundance} illustrates our results, including the scaling of $\simpss$  with the migration probability $\mig$ and a plot of three steady-state distributions.

\subsection{Steady-state Simpson index}
\label{sec:ss-simps}

In this section, we find the Simpson index as a function of $\mig$ and $N$. 
We begin by writing $\simpss$ in terms of the species abundance distribution $P_n$ as  
\begin{align}
\simpss\approx 
\frac{1}{N}\frac{\sum_{n=1}^N P_n n^2}{\sum_{n=1}^N P_n n}\,.
\label{eq:simpPn}
\end{align}
This expression, derived in SM\ref{app:ss-diversity}, implies that the expected Simpson index is proportional to the ratio of the second and first moments of $P_n$. 
Since $P_n$ itself depends on $\simpss$,
we must solve \eqref{eq:simpPn} using self-consistency. We do this separately in the low-migration and the high-migration regimes.  
We also consider several important transition points.

\subsubsection{Low migration} 
\label{sec:low-mig}

As shown in the SM\ref{app:ss-diversity}, in the low-migration regime, we may approximate the two sums in \eqref{eq:simpPn} by integrals. This leads to the equation
\begin{align}
    \frac{1}{\mig} = \sqrt{\frac{\pi N}{2}}\erfcx\left(\sqrt{\frac{N}{2}}(\mig-\simpss)\right)\,,
    \label{eq:solA0}
\end{align}
where $\erfcx(z):=e^{z^2}\text{erfc}(z)=e^{z^2}\sqrt{4/\pi}\int_{z}^{\infty}e^{-t^{2}}\mathrm{d}t$ is the \emph{scaled complementary error function}~\cite{zaghloul2024efficient}. 
Although the function $\erfcx(z)$ does not have a closed-form inverse, it has an efficient numerical implementation~\cite{johnson2012faddeeva} that allows us to quickly compute $\simpss$ using numerical root-finding algorithms.  
The scaling predicted by solving this equation for fixed $N$ and varying $\mig$ is shown as a red curve in Fig.~\ref{fig:speciesabundance}(a). 

A closed-form approximation is possible for sufficiently small $\mig$. For $\mig-\simpss \ll 0$, we may use $\erfcx(z)\approx 2e^{z^2}$ for $z\to -\infty$. Plugging into \eqref{eq:solA0} and solving gives 
\begin{align}
\simpss \approx \mig + \sqrt{\frac{-\ln\left(2\pi N \mig^2\right)}{N}}\approx \sqrt{\frac{-\ln\left(2\pi N \mig^2\right)}{N}}.\label{eq:ss-div}
\end{align}
Note that this approximation is only defined for $\mig \le 1/\sqrt{2 \pi N}$, so that the argument of the square root is nonnegative. 

\eqref{eq:ss-div} implies that, in the low-migration regime, the Simpson index scales as $\sim 1/\sqrt{N}$
in population size and as $\sim\sqrt{-\ln \mig}$ in the migration probability. The scaling predicted by \eqref{eq:ss-div} is shown as a blue curve in Fig.~\ref{fig:speciesabundance}(a). The predicted abundance distribution is compared against simulations in 
Fig.~\ref{fig:speciesabundance}(b). 

Interestingly, an expression similar to Eq.~\eqref{eq:solA0} has previously appeared in a different ecological model, which considers a population embedded in a fluctuating spatiotemporal fitness landscape~\cite[Appendix~C]{lofflerRandomFitness2020}. 

\subsubsection{Bimodality point}

We now find the migration probability below which the species abundance distribution becomes bimodal.  
To do so, we consider $P_n$~\eqref{eq:Pn1} as a differentiable function of $n$ and find the critical abundances $\hat{n}$ where its derivative vanishes. With some algebra, these are found to be
\begin{align}
    \hat{n}=\frac{N}{2} \Big[\simpss + \ln(1-\mig) \pm \sqrt{(\simpss +  
\ln(1 - \mig))^2-4/N}\Big]\,.
\label{eq:hatn}
\end{align}
We simplify by expanding around large $N$ and small $\mig$ to express the two critical abundances as
\begin{align}
    \hat{n}_{\min} \approx \frac{1}{\simpss}\quad \text{and}\quad \hat{n}_{\max} \approx N \simpss - \frac{1}{\simpss},
    \label{eq:nminmax}
\end{align}
shown as the local minimum and local maximum in Fig.~\ref{fig:schematicss}. These two points become distinct when
\begin{align}
    \simpss \ge \frac{2}{\sqrt{N}} \quad\text{and}\quad \mig \le \mig_B := \frac{e^{-2}}{\sqrt{2 \pi N}}\,,
    \label{eq:bimodality}
\end{align} 
where $\mig_B$ is found by plugging $ \simpss = 2/\sqrt{N}$ into \eqref{eq:ss-div} and solving. Thus,  $\mig_B$ is the migration probability below which we see the formation of a bimodal abundance distribution.

\subsubsection{Logseries onset}

Another interesting value for the 
migration probability 
separates the cooperation- from the migration-dominated regimes. As we increase $\mig$ towards 1, the abundance distribution becomes dominated by the Logseries term, recovering the distribution predicted by Hubbell's neutral theory of biodiversity.

To identify this transition point, we consider the migration rate at which $\mig$ becomes larger than $\simpss$. Plugging $\mig-\simpss=0$ into \eqref{eq:solA0} and using $\erfcx(0)=1$ specifies this point as 
\begin{align}
\mig_L:=\sqrt{\frac{2}{N\pi}}\,.\label{eq:transition}
\end{align}

\subsubsection{High migration}

As mentioned above, when $\mig$ is large, 
the abundance distribution approaches that of Hubbell's neutral model. 
To estimate the steady-state Simpson index $\simpss$ in this regime, observe that the term $({1-\mig})^n$ in \eqref{eq:Pn1} imposes an exponential cutoff on abundances $n >1/\mig$. Considering the regime $\mig > \mig_L$ from \eqref{eq:transition}, we may restrict our attention to $n < 1/\mig < \sqrt{N\pi/2}$ and $\simpss < \mig$. For these values, the Gaussian term can be approximated as a constant factor, $e^{-(n-N\simpss)^2/2N}\approx e^{-N\simpss^2/2}$. 
We re-evaluate $\simpss$~\eqref{eq:simpPn} 
using the Logseries distribution $P_n \propto ({1-\mig})^n/n$ and large $N$, giving
\begin{align}
    \simpss \approx  \frac{1}{N \mig}\,.
    \label{eq:simphighmig}
\end{align}

Thus, with high migration, the Simpson index scales as $\simpss\sim N^{-1}$
in population size and $\simpss\sim \mig^{-1}$ in migration probability. 
This scaling is shown with the orange curve in Fig.~\ref{fig:speciesabundance}(a).

In fact, as shown in SM\ref{app:high-migration-simpss-from-erfcx}, 
the high-migration expression~\eqref{eq:simphighmig} can be derived from \eqref{eq:solA0}. This is surprising because \eqref{eq:solA0} was derived under the assumption of $\mig \ll \simp$. Empirically, we found that the value of $\simpss$ given by the numerical solution to \eqref{eq:solA0} provides an excellent match to $\simpss$ calculated from simulations, even for large migration probabilities (see Fig.~\ref{fig:speciesabundance}(a), red curve).

\begin{figure}[t]
\begin{center}
\includegraphics[width=\columnwidth]{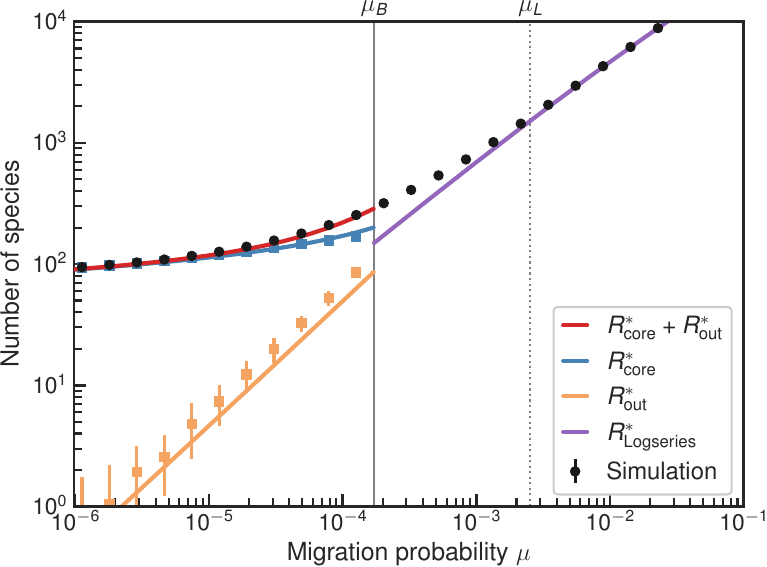}
\end{center}
\caption{\textbf{Scaling of number of species $R^*$ versus migration probability $\mig$} ($N=10^5$). We compare mean and standard deviations from simulations of the full system (100 runs; black); approximations in the low-migration regime of the number of total species (red), core species~\eqref{eq:Rcore_estimate} (blue), and non-core species~\eqref{eq:Rout_estimate} (orange);  approximations in the high-migration regime of the total number of species (purple). 
Solid vertical line indicates $\mu_B$~\eqref{eq:bimodality} where bimodality is lost; dotted vertical line indicates  $\mu_L$~\eqref{eq:transition} where system transitions to Logseries regime.
}
\label{fig:numspecies}
\end{figure}

\subsection{Steady-state number of species}

As another important measure of diversity, we consider the expected number of different species in steady state:
\begin{align}
    R^*:=\langle R(\n)\rangle\ .
\end{align}
The expected steady-state abundance of the representative species obeys 
$\sum_n P_n n = N/R^*$, therefore the number of species can be found as
\begin{align}
    R^* =\frac{N}{\sum_n P_n n}\,.\label{eq:Rss}
\end{align}

With high migration, $R^*$ can be estimated by evaluating the denominator of \eqref{eq:Rss}. We may ignore the Gaussian contribution in \eqref{eq:Pn1}, since the most relevant contribution to \eqref{eq:Pn1} is the Logseries distribution. For large $N$, the normalization constant of the Logseries term is $-\ln\mig$, thus we compute 
\begin{align}
    \sum_n P_n n\approx\frac{1}{-\ln\mig}\sum_n (1-\mig)^n\approx \frac{1-\mig}{-\mig\ln \mig} .
    \label{eq:fisherR}
\end{align}
Substituting back into \eqref{eq:Rss} gives
\begin{align}
    R^*\approx\RFisher:= -\frac{\mig N}{1-\mig}\ln\mig,\label{eq:Rss_outside_bimodality}
\end{align}
shown in purple in Fig.~\ref{fig:numspecies}. As expected, in the limit of $\mig \to 1$, $\RFisher\to N$. This indicates that if only migration occurs (with no replication taking place), then the system acquires the maximum number of species with each of the $N$ individuals belonging to a different species.

In the low-migration regime, below the point of bimodality~\eqref{eq:bimodality},  there are two relevant species counts.
The first is $\Rcore$, the number of high-abundance species that belong to the cooperator core. We define these species as those with abundances larger than the local minimum, $n > 1/\simpss$, see Fig.~\ref{fig:schematicss}.
The second is $\Rout$, the number of low-abundance species that remain {outside} of the cooperator core and have abundance $n \le 1/\simpss$. 
The total number of species is given by
\begin{align}
    R^*\approx \Rcore + \Rout\,. \label{eq:Rss_bimodal}
\end{align}

Let us introduce $N_{\textsf{core}}$ and $N_{\textsf{out}}$ as the total number of individuals inside and outside the core, respectively, where $N_{\textsf{core}}+N_{\textsf{out}}=N$. We also denote by $\langle n \rangle_{\textsf{core}}$ and $\langle n \rangle_{\textsf{out}}$ the expected abundances of a species conditioned to be inside and outside the core, respectively. 
To estimate $\Rcore$, we use the relation $\Rcore = N_{\textsf{core}}/\langle n \rangle_{\textsf{core}}$. As shown in  SM\ref{app:num_species}, to a first approximation we may take $N_{\textsf{core}}\approx N$ (i.e., in the low-migration regime, most individuals belong to the core). Then, using $\langle n\rangle_{\textsf{core}}\approx\hat{n}_{\textsf{max}}$ from \eqref{eq:nminmax}, we obtain
\begin{align}
\Rcore \approx \frac{1}{\simpss-(N\simpss)^{-1}},\label{eq:Rcore_estimate}
\end{align}
shown in blue in Fig.~\ref{fig:numspecies}.
Similarly, we estimate $\Rout$ by using $\Rout=N_{\textsf{out}}/\langle n\rangle_{\textsf{out}}$. We may estimate this ratio as (see  SM\ref{app:num_species}):
\begin{align}
    \Rout \approx -N\mig\ln \simpss, \label{eq:Rout_estimate}
\end{align}
shown in orange in Fig.~\ref{fig:numspecies}. The red curve shows the total number of species $R^*\approx \Rcore + \Rout$. Observe that at low migration $\mig$, almost all species belong to the core.  

\begin{figure}[t]
\begin{center}
\includegraphics[width=\columnwidth]{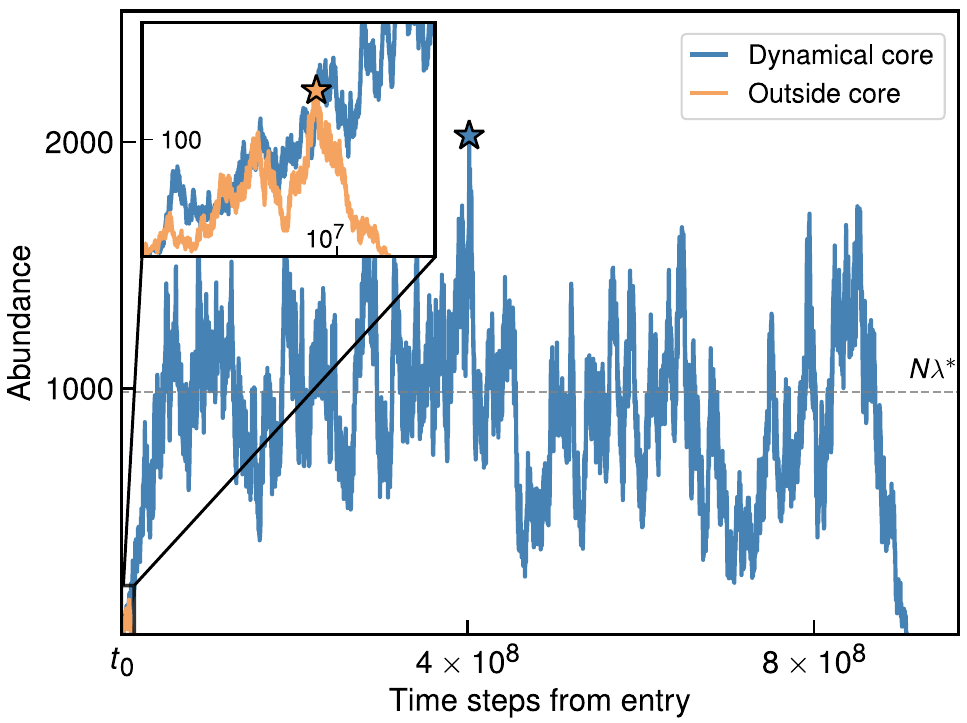}
\caption{\textbf{Abundance trajectories from entry to extinction}, illustrating that species that enter the dynamical core reside for much longer times. We sample two abundance trajectories from the stationary process ($N=10^5,\mig=10^{-5}$), one for a species 
that enters the dynamical core and one for a species that does not. 
Dashed gray line indicates the abundance position of the local minimum discussed in Sec.~\ref{sec:dyn}~\ref{sec:infiltration-probability}, stars indicate maximum abundances reached. 
\label{fig:trajectories}
}
\end{center}
\end{figure}

\begin{figure}[t]
\begin{center}
\includegraphics[width=\columnwidth]{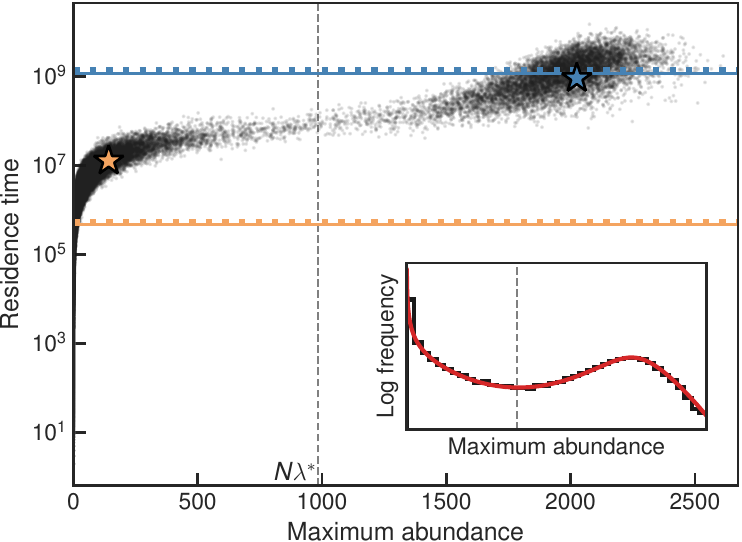}
\caption{\textbf{Species residence times and maximum abundances} for $10^6$ species sampled from stationary dynamics ($N=10^5$, $\mig=10^{-5}$). The species fall into two clusters: one with long residence times and high maximum abundances, the \emph{dynamical core}, and one with short residence times and low maximum abundances. The residence times and maximum abundances of the two trajectories from Fig.~\ref{fig:trajectories} are indicated with star symbols. 
Horizontal lines indicate predicted (solid) and empirical (dashed) mean residence times of species in the dynamical core (blue) and outside of the dynamical core (orange). Inset: the empirical histogram (black) and prediction (red, from Eqs.~\eqref{eq:QnUn} and~\eqref{eq:u1_inv_n0}) of maximum abundances distribution $Q_m$. The distribution has a bimodal shape, with the dividing point located near abundance $N\simpss$ (dashed vertical line in main plot and inset). 
\label{fig:scatter}
}
\end{center}
\end{figure}

\mainsection{Dynamical properties}
\label{sec:dyn}

We now study the 
stochastic dynamics of our model by considering trajectories of species as they enter and leave the population. We focus mainly on the low-migration regime, where cooperation plays an important role. 

The dynamics of any given species is described by a trajectory of nonnegative abundance values $n(t)$, starting from $n(t_0)=1$ when that species enters the population at time $t=t_0$ and ending on $n(t_f)=0$ when that species goes extinct at time $t=t_f$. 
We characterize each trajectory by two statistics: the maximum abundance reached, $\max_{t:t_0\le t\le t_f} n(t)$, and the residence time before extinction, $t_f - t_0$.  

For concreteness, we illustrate two typical abundance trajectories in Fig.~\ref{fig:trajectories}. We see that the blue trajectory resides in the system for a very long time and reaches a high maximum abundance value, while the orange trajectory does not reach a high abundance and quickly goes extinct. 

In fact, simulations show that all trajectories cluster into two well-defined classes: one with long residence times and high abundances, which we term the \emph{dynamical core}, and one with short residence times and low abundances. 
To illustrate this, Fig.~\ref{fig:scatter}, shows a scatter plot of the maximum abundances and residence times of $10^6$ randomly sampled species. Species that fall into the dynamical core (top right cluster) tend to have residence times orders of magnitude larger than those that do not enter the dynamical core. Moreover, the distribution of maximum abundances has a clear bimodal shape (inset), allowing us to define a quantitative threshold for classifying species as belonging to the dynamical core.

In the following, we derive the distribution of maximum abundances, shown in red in Fig.~\ref{fig:scatter}. We then use this distribution to define the abundance threshold for the dynamical core. We also derive the mean residence times of species that belong and do not belong to the dynamical core.

\subsection{Maximum abundances}
\label{sec:infiltration-probability}

Let $Q_m$ indicate the probability that a typical species trajectory reaches a maximum abundance $m$. To find the form of $Q_n$, it will be helpful to introduce the quantity $u_n(\ell)$ to represent the probability that a species with abundance $n$ reaches abundance $\ell$ or higher before going extinct. $Q_m$ and $u_n(\ell)$ are related by 
\begin{align}
Q_m=u_{1}(m)-u_1(m+1)\,.\label{eq:QnUn}
\end{align}
In other words, $Q_m$ is equal to the probability that a species that enters the system at abundance $n=1$ eventually reaches abundance $m$ but not abundance $m+1$.

Importantly, the birth-death process defined by~\eqref{eq:birthprob}-\eqref{eq:deathprob}, $u_n(\ell)$ obeys the following recurrence:
\begin{align}
    u_n(\ell) = \forward{n} u_{n+1}(\ell)+\backward{n}u_{n-1}(\ell)+(1-\forward{n}-\backward{n})u_n(\ell).\label{eq:inf_prob_recurrence}
\end{align}
The boundary conditions are $u_0(\ell)=0$ (extinct species never come back into the system) 
and $u_\ell(\ell)=1$. A general solution to the recurrence~\eqref{eq:inf_prob_recurrence} is derived in SM\ref{app:infiltration-prob}. For $u_1(\ell)$, the quantity that enters into \eqref{eq:QnUn}, this solution is approximated as
\begin{equation}
\scalebox{0.8725}{$
u_1(\ell) \approx \left[
e^{\ell^2/2N - (\ell - 1)\simpss}
\left( \sqrt{2N}\DawsonOrig(x_{\ell}) + \frac{1}{2} \right)
- \sqrt{2N}\DawsonOrig(x_1) + \frac{1}{2}
\right]^{-1},
$}\label{eq:u1_inv_n0}
\end{equation}
where  we introduce the  rescaled abundances $x_{\ell} := (\ell - N\simpss)/{\sqrt{2N}}$ and the Dawson function 
$\DawsonOrig(z):=e^{-z^2}\intop_0^ze^{t^2}dt$. 

Plugging this result into \eqref{eq:QnUn} gives the probability  $Q_m$ that a species reaches maximum abundance $m$. The predicted and empirical distribution of maximum abundances is shown in Fig.~\ref{fig:scatter} (inset). 
In addition, the distribution $Q_m$ is bimodal, with a local minimum located at approximately $N\simpss$ (see  SM\ref{app:infiltration-prob}). 
This bimodality 
allows us to classify species into two sets: those whose abundances reach $N\simpss$ (or higher) and those that do not.  We refer to the former as the \emph{dynamical core}. 
For example, Fig.~\ref{fig:trajectories} shows a blue trajectory that reaches abundance values larger than $N\simpss$  (dashed line) and thus belongs to the dynamical core, while the orange trajectory does not reach this threshold and thus remains outside the dynamical core.

Finally, we use the term \emph{infiltration probability} $\beta$ to refer to the probability that a new species will enter the dynamical core. 
Interestingly, as we show in SM\ref{app:infiltration-prob}, this probability is approximately equal to the Simpson index, 
\begin{align}
\beta \approx \simpss,\label{eq:beta}
\end{align}
suggesting that 
new species are less likely to enter the dynamical core in more diverse populations.

\subsection{Residence times}
We now derive the mean residence times by separately considering species that enter the dynamical core ($\mrtcore$), that do not enter the dynamical core ($\mrtout$), and all species ($\mrt$). 
For simplicity, we will also use our calculations of the expected number of all present species $R^*$~\eqref{eq:Rss_outside_bimodality}-\eqref{eq:Rss_bimodal}, core species $\Rcore$~\eqref{eq:Rcore_estimate}, and non-core species $\Rout$~\eqref{eq:Rout_estimate}. 

Here, there is a subtle point to be raised. Our calculations of $\Rcore$ and $\Rout$ were based on a ``static'' definition of the core, as the set of species that have abundance greater or smaller than $1/\simpss$ at a single point in time. This is different from the ``dynamic'' definition of the core considered in this section, as the set of species whose trajectory reaches an abundance greater than $N\simpss$ at any point in time. 
In SM\ref{app:mrt}, we show that mean residence times can be alternatively derived using a mean first-passage time (MFPT) calculation that does not explicitly invoke Eqs.~\eqref{eq:Rss_outside_bimodality}-\eqref{eq:Rout_estimate}. This alternative MFPT-based analysis, which we do not include here for simplicity, leads to the same quantitative results. This suggests that the two definitions of the core are essentially equivalent when restricted to the set of species present in steady state.

\begin{figure}[t]
\begin{center}
\includegraphics[width=\columnwidth]{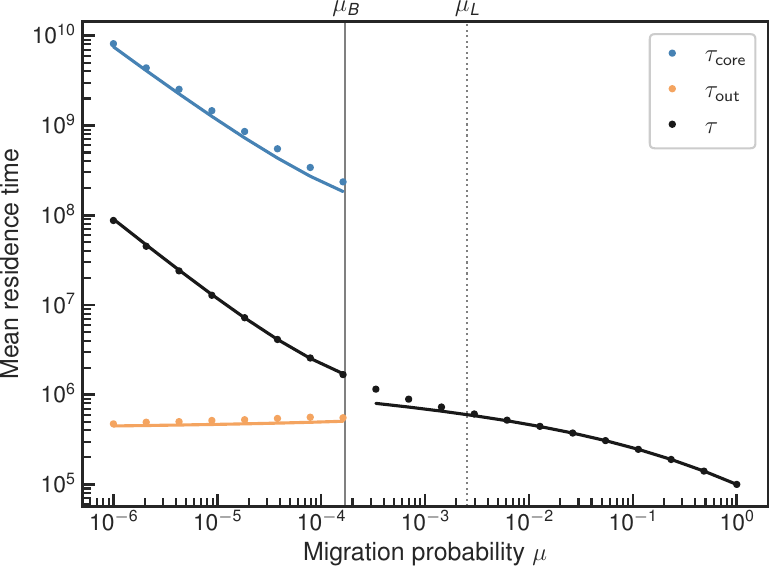}
\caption{\textbf{Scaling of mean residence times} with migration probability $\mig$ ($N=10^5$). Simulations (dots) are compared to predictions (solid curves) for: core species, $\mrtcore$~\eqref{eq:mrtcore}; non-core species, $\mrtout$~\eqref{eq:mrtout}; and all species, $\mrt$~\eqref{eq:mrtall}-\eqref{eq:av_entry-exit_rates} for low and high values of $\mig$. Solid vertical line indicates $\mu_B$~\eqref{eq:bimodality} where bimodality is lost; dotted vertical line indicates  $\mu_L$~\eqref{eq:transition} where system transitions to Logseries regime.
\label{fig:residencetimes}
}
\end{center}
\end{figure}

We now consider the low-migration regime, and derive the mean residence times of species that enter the 
core, $\mrtcore$, and those that do not enter the core, $\mrtout$.
The rate at which new species migrate into the system and eventually enter the dynamical core is given by $\mig\beta$. On the other hand, the rate at which core species go extinct is $\Rcore/\mrtcore$, where $\Rcore$ is the expected number of species in the core at any one time. 
Since these rates must balance in steady state, we have  $\mig\beta=\Rcore/\mrtcore$. Using 
Eqs.~\eqref{eq:Rcore_estimate} and \eqref{eq:beta}, we arrive at 
\begin{align}
\mrtcore\approx \frac{1}{\mig\simpss\left[\simpss-(N\simpss)^{-1}\right]}.
\label{eq:mrtcore}
\end{align}
Similarly, for species that never reach the core, entry rates must be balanced against extinction rates as $\mig(1-\beta)=\Rout/\mrtout$. Using Eqs.~\eqref{eq:Rout_estimate} and \eqref{eq:beta}, this gives
\begin{align}
    \mrtout \approx -\frac{N}{1-\simpss}\ln \simpss
\label{eq:mrtout}
\end{align}
We estimate the overall mean residence time by combining these results with the infiltration probability as 
\begin{align}
    \mrt \approx 
    \beta\mrtcore+(1-\beta)\mrtout.
    \label{eq:mrtall}
\end{align}
These predictions are compared against data in Fig.~\ref{fig:scatter} (solid and dotted horizontal lines). 

The scaling of mean residence times against migration probability $\mig$ is shown in 
Fig.~\ref{fig:residencetimes}. 
For low migration probabilities, core species live orders of magnitude longer than species that do not enter the core.

Our analysis above has mostly focused on the low-migration regime, in which the distinction between core and non-core species is meaningful. In the high-migration regime, we consider the mean residence time of all species $\tau$. In steady state, 
entry and exit rates of all species into the system must be balanced, $\mu = {R^*}/{\mrt}$. Using our estimate of $R^*$\eqref{eq:Rss_outside_bimodality} gives
\begin{align}
\mrt \approx 
\frac{-N}{1-\mig}\ln \mig.\label{eq:av_entry-exit_rates}
\end{align}

\mainsection{Discussion}
\label{sec:discussion}

In this paper, we introduced a neutral model for cooperative ecosystems, deliberately choosing the simplest set of rules consistent with cooperative interactions. 
This minimalist design, inspired by the neutral theory of biodiversity~\cite{hubbell1997unified}, allows a full analytical treatment while capturing the essential features of cooperative dynamics. Our analysis addresses both steady-state behavior and the dynamical properties of the system, providing a basic baseline for future comparisons with more complex models. It also complements existing models of cooperation, such as 
mutualistic voter models~\cite{tu2019reconciling}, where agents adopt neighbor states with a bias toward cooperation, or mutualistic Lotka-Volterra systems~\cite{Holland2005}, where species benefit each other's growth.

In the first part, we derived an expression for the steady-state Simpson diversity index, which allowed us to classify the system into different regimes depending on the migration probability. At high migration rates,  species abundances are very low and frequency-dependent effects become negligible. In this regime, we recover the predictions of Hubbell's neutral theory of biodiversity, including a Logseries species abundance distribution in steady state. 
In contrast, at low migration rates, our model exhibits frequency-dependent reproduction rates that result from cooperative interactions. In this regime, our model predicts the emergence of a bimodal abundance distribution (see Fig.~\ref{fig:schematicss}), which cannot be derived from the classic neutral theory.

Bimodality allows us to define a core of cooperators, defined as the set of species that belong to the high-abundance component. We derive scaling laws for the core, showing that the effective number of species scales roughly as $N^{1/2}$ with population size and $(-\ln \mig)^{-1/2}$ with migration rate. Due to the presence of the core, the system preserves diversity (maintains low $\simpss$) even at exponentially small migration rates. Hence, cooperative interactions can dramatically increase ecosystem stability.

Several studies have reported bimodal abundance distributions in gut microbiomes; see, for example, Refs.~\citep[][Fig.~1]{loftus2021bacterial} and~\citep[][Fig.~2]{costea2018enterotypes}. 
Previous research has attributed bimodality to emergent niche partitioning~\cite{vergnon2012emergent}, intrinsic bistability~\cite{gonze2017multi}, or a combination of multiple processes~\cite{costea2018enterotypes}. However, our theory suggests a neutral mechanism that generates bimodality under very minimal assumptions, offering a novel explanation of this phenomenon.

In the second part of our paper, we studied the stochastic dynamics of species as they enter and leave the system. In the low-migration regime, the distribution of maximum abundances again exhibits a bimodal shape, allowing us to classify species into two types. This classification captures the dynamical signature of the cooperator core, as discussed in the steady-state analysis in the first part of this paper. We showed that species that enter the core achieve much higher abundances and longer residence times than species that do not enter the core.

It is worth noting that other models involving higher-order interactions have been proposed and investigated in various contexts. 
For example, Hinrichsen~\cite{hinrichsen2001pair} investigated the universality class of the pair contact process $2A\to 3A$, $2A\to 0$ with diffusion and suggested that this model does not belong to the directed percolation universality class.  In Ref.~\cite{bessonov2017phase}, the authors studied a quadratic contact process and determined a bound on the critical value of the infection rate, while in Refs.~\cite{iacopini2019simplicial} and~\cite{kim2024higher}, the authors introduced general higher-order social contagion models and showed that these higher-order interactions fundamentally change the nature of the phase transition to a contagion phase. Although these models are related to ours in their use of higher-order interactions, they do not capture key features of our system, such as the bimodal species abundance distribution and the emergence of a cooperator core. Moreover, other similarities can be found when comparing our work with a recent Generalized Lotka-Volterra model with annealed disorder, which shows how temporal fluctuations in interactions act as effective environmental noise, promoting diversity~\cite{suweis2024generalized}. Though based on different mechanisms, both highlight how non-equilibrium stochastic dynamics can generate high diversity and bimodal species abundance patterns.

In future work, it may be possible to generalize the method developed in this paper to other models of neutral cooperators, including three-way generalizations of our pairwise cooperation rule as well as other models considered in the literature. For a general neutral model, one may describe the abundance fluctuations of a ``representative species'' using a nonlinear birth-death master equation~\cite{malek-mansour_master_1975}, with transition rates encoding mean-field effects of the rest of the population. Steady-state properties can then be identified by solving with self-consistency, in a similar way as done in this study.

Another promising direction to explore is the use of spatially extended models, which can offer additional insights into the effects of local interactions. This approach has also been extensively developed within the framework of neutral theory~\cite{hubbell1997unified,etienne2007neutral,o2010field}, where it has enabled the study of the dynamics of metacommunities. In our cooperative context, spatial structure would introduce correlations between neighboring organisms, effectively reducing the reproduction rate of existing species. We anticipate that this rate would scale with the surface area that separates a given species from its neighbors. Consequently, less frequent species—having a higher surface-area-to-volume ratio—would gain a relative advantage over more common species, which have a lower ratio. As a result, we expect that the mean abundance of the dominant core species would decrease in spatially structured environments.

Experimental validation of our predictions could be achieved using engineered microbial consortia. For instance, a synthetic community could be constructed in which each strain lacks the ability to synthesize a single essential metabolite but can obtain it through cross-feeding with the others. Such syntrophic systems have been successfully engineered, ranging from two-species~\cite{shou2007synthetic,amor2017spatial} to multispecies communities~\cite{mee2014syntrophic}. This approach would allow us to test our theoretical predictions under relaxed connectivity constraints, extending beyond the fully cooperative scenario assumed here. Minimal genetic modifications could be introduced into a single model organism (such as E. coli) to generate multiple auxotrophic variants that remain metabolically and physiologically equivalent, thus approximating a neutral interaction regime. By labeling some of these variants with a distinct fluorescent reporter, one could perform time-resolved sampling and quantify strain-specific abundance distributions using flow cytometry.

Further extensions of the theory should consider the heterogeneity of interactions, complex network topologies, and spatial effects. Another interesting direction would be to build a connection to the theory of autocatalytic chemical reaction networks, as studied in research on the origin of life. In particular, our approach is related to the ``hypercycle''~\cite{eigen2012hypercycle,szathmary1987group,schuster2016some,sole2025bifurcations}, a proposed model of collective chemical replication. In this model, a directed network of chemical species ($X_i$) cross-catalyze the replication of other species ($X_i + X_j \to X_i +X_i + X_j$ for neighbors $i \to j$).  This reaction scheme is equivalent to our replication rule, Fig.~\ref{fig:1}, therefore our model can be interpreted as a (stochastic) hypercycle on a fully-connected network.  In this context, our analysis could shed light on the emergence of cross-catalytic cores in replicating chemical systems.

\vspace{10pt}

\begin{acknowledgments}
JP and RS thank the hospitality of the Santa Fe Institute, where this work started. JP also thanks the hospitality of the Simon Levin Lab where some of this work was developed. RS acknowledges support from the Institució Catalana de Recerca i Estudis Avançats
(ICREA), the AGAUR 2021 SGR 0075 grant, a PID2023-152129NB-I00 grant funded by MICIU/AEI/10.13039/501100011033 and by the Santa Fe Institute. This project was partly supported by Grant No. 62417 from the John Templeton Foundation. The opinions expressed in this publication are those of the authors and do not necessarily reflect the views of the John Templeton Foundation. AK was partly supported by the European Union’s Horizon 2020 research and innovation programme under the Marie Skłodowska-Curie Grant Agreement No.~101068029. Finally, SR thanks NSF grant DMR-1910736 for financial support.

\end{acknowledgments}

\vfill
\bibliographystyle{ieeetr}
\bibliography{main}

\clearpage

\newcommand\newstuff[1]{#1}

\ifarxiv 
\let\addcontentsline\savedaddcontentsline
\input{sm-contents}
\fi

\end{document}